\documentclass[11pt,letterpaper,twoside]{article}
\usepackage[centertags]{amsmath}
\usepackage{graphicx,indentfirst,amsmath,amsfonts,amssymb,amsthm,newlfont}
\font\Goth=yinitas scaled \magstep0

\newcommand{\Gth}[1]{\lower2mm\hbox{\Goth #1}}

\usepackage[latin1]{inputenc}
\def\de{\delta}

\def\l1{{\lambda}_1}

\newcommand{\f}{\frac}

\def\ln{\mbox{\rm ln}}

\def\x1{{\xi }_{xx}}
\def\x2{{\xi }_{yy}}
\def\x3{{\xi }_{xy}}

\def\e1{{\eta }_{xx}}
\def\e2{{\eta }_{yy}}
\def\e3{{\eta }_{xy}}

\newcommand{\ds}{\displaystyle }
\newtheorem{definition}{Definition}
\newtheorem{theorem}{Theorem}

\newcommand{\beqn}{\begin{eqnarray*}}
\newcommand{\eeqn}{\end{eqnarray*}}
\newcommand{\beqnn}{\begin{eqnarray}}
\newcommand{\eeqnn}{\end{eqnarray}}
\newcommand{\p}{\partial}
\newcommand{\bb}{\begin{equation}}
\newcommand{\ee}{\end{equation}}
\newcommand{\ba}{\begin{array}}
\newcommand{\ea}{\end{array}}

\begin{document}
\pagenumbering{arabic}
\title{\huge \bf New classes of nonlinearly self-adjoint evolution equations of third- and fifth-order}
\author{\rm \large Igor Leite Freire\\
\\
\it Centro de Matemática, Computação e Cognição\\ \it Universidade Federal do ABC - UFABC\\ \it 
Rua Santa Adélia, $166$, Bairro Bangu,
$09.210-170$\\\it Santo André, SP - Brasil\\
\rm E-mail: igor.freire@ufabc.edu.br/igor.leite.freire@gmail.com}
\date{\ }
\maketitle
\vspace{1cm}
\begin{abstract}
In a recent communication Nail Ibragimov introduced the concept of nonlinearly self-adjoint differential equation [N. H. Ibragimov, Nonlinear self-adjointness and conservation laws, J. Phys. A: Math. Theor., vol. 44, 432002, 8 pp., (2011)]. In the present communication a nonlinear self-adjoint classification of a general class of fifth-order evolution equation with time dependent coefficients is presented. As a result five subclasses of nonlinearly self-adjoint equations of fifth-order and four subclasses of nonlinearly self-adjoint equations of third-order are obtained. From the Ibragimov's theorem on conservation laws [N. H. Ibragimov, A new conservation theorem, J. Math. Anal. Appl., vol. 333, 311--328, (2007)] conservation laws for some of these equations are established.
\end{abstract}
\vskip 1cm
\begin{center}
{2000 AMS Mathematics Classification numbers:\vspace{0.2cm}\\
76M60, 58J70, 35A30, 70G65\vspace{0.2cm} }
\end{center}
\pagenumbering{arabic}
\newpage

\section{Introduction}

Probably the most famous result connecting symmetries of differential equations and conservation laws is the well known Noether theorem, which allows one to construct conservation laws for a differential equation following a straightforward algorithm.

Although Noether's approach provides an elegant algorithm for finding conservation laws, it possesses a strong limitation: it can only be applied to equations having variational structure. However a large number of differential equations without variational structure admits consevation laws. A trivial example is the heat conduction equation 
$$D_{t}(u)-D_{x}(u_{x})=u_{t}-u_{xx}=0,$$
which is itself a conservation law.

It has been subject of intense research to find methods for constructing conservation laws for equations without variational structure. In this direction Nail Ibragimov proved a result \cite{ib2} recently, which we would like to refer as Ibragimov's theorem on conservation laws.

Let $x=(x^{1},\cdots,x^{n})$ be $n$ independent variables, $u=u(x)$ be a dependent variable,
\bb\label{1.1}
X=\xi^{i}\f{\p}{\p x^{i}}+\eta\f{\p}{\p u}+\eta_{i}\f{\p}{\p u_{i}}+\eta_{ij}\f{\p}{\p u_{ij}}+\cdots,
\ee
be a symmetry of an equation 
\bb\label{1.2}
F(x,u,\cdots,u_{(s)})=0
\ee
and
\bb\label{1.2'}
\f{\de}{\de u}=\f{\p}{\p u}+\sum_{j=1}^{\infty}(-1)^{j}D_{i_{1}}\cdots D_{i_{j}}\f{\p}{\p u_{i_{1}\cdots i_{j}}},
\ee
be the Euler-Lagrange operator.

The Ibragimov's theorem on conservation laws states the following.

\begin{theorem}\label{new}
Let $(\ref{1.1})$ be any symmetry (Lie point, Lie-Bäcklund, nonlocal symmetry) of equation $(\ref{1.2})$,
\bb\label{1.3}
F^{\ast}(x,u,v,\cdots,u_{(s)},v_{(s)}):=\f{\de}{\de u}{\cal L}=0,
\ee
where ${\cal L}=vF$ is the formal Lagrangean, be the adjoint equation to equation $(\ref{1.2})$. The combined system $(\ref{1.2})$ and $(\ref{1.3})$ has the conservation law $D_{i}C^{i}=0$, where
\bb\label{1.4}
\ba{lcl}
C^{i}&=&\ds{\xi^{i}{\cal L}+W\,\left[\f{\p{\cal L}}{\p u_{i}}-D_{j}\left(\f{\p{\cal L}}{\p u_{ij}}\right)+D_{j}D_{k}\f{\p{\cal L}}{\p u_{ijk}}-\cdots\right]}\\
\\
&&\ds{+D_{j}(W)\,\left[\f{\p{\cal L}}{\p u_{ij}}-D_{k}\left(\f{\p{\cal L}}{\p u_{ijk}}\right)+\cdots\right]}\\
\\
&&\ds{+D_{j}D_{k}(W)\,\left[\f{\p{\cal L}}{\p u_{ijk}}-\cdots\right]+\cdots}
\ea
\ee
and $W=\eta-\xi^{i}u_{i}$.
\end{theorem}

On one hand note that the Ibragimov's theorem provides a so elegant algorithm for finding conserved vectors as the Noether's approach provides. On the other hand it is important to observe that the Ibragimov's theorem on conservation laws does not require the existence of a Lagrangean and it can be applied for any differential equation. Indeed it can be extended to systems of differential equations, see the prominent paper \cite{ib1}.

Although its generality, the Ibragimov's theorem on conservation laws, {\it a priori}, provides a nonlocal conservation law for the original equation, that is, the components $C^{i}$ of the conserved vector depend on the variables $u$ and $v$, not only $u$. It is the price for using this powerful and general result. However, for a some special classes of equations it is possible to find local conservation laws from Ibragimov's theorem.

In \cite{ib1,ib2} Ibragimov introduced the concept of {\it self-adjoint differential equations}. Later Ibragimov extended this concept to quasi-self-adjoint differential equations, see \cite{ib3}. Recently both concepts were generalized to weak self-adjoint differential equations by Maria Luz Gandarias \cite{gandjpa} and to nonlinearly self-adjoint differential equations by Nail Ibragimov \cite{ib6,ib8}. 

These special equations have the remarkable property to make the adjoint equation equivalent to the original one upon an appropriated substitution. Then the conserved vector, with such substitution, becomes a local conserved vector, providing a local conservation law for the original equation. These concepts will be discussed in the section \ref{non}. 

Since Ibragimov's seminal works \cite{ib1,ib2,ib3} intense research has been carried out in order to find self-adjoint and quasi-self-adjoint classes of equations and, by using theorem \ref{new}, conservation laws for these equations have been established. Most of these works have been focused in evolution equations.

Namely Ibragimov, Torrisi and Tracinà determined that the system derived from a $(2+1)$ generalized Burgers equation is quasi-self-adjoint in \cite{ib4}. Bruzón, Gandarias and Ibragimov determined necessary and sufficient conditions for a general fourth-order evolution equation to be self-adjoint in \cite{b1}. Quasi-self-adjointness of a generalization of the Camassa-Holm equations was obtained by Ibragimov, Khamitova and Valenti in \cite{ib7}. In \cite{tori} a quasi self-adjointness classification of quasilinear dispersive equations was carried out by Torrisi and Tracinà. Further examples can be found in \cite{ijnmp,iaaca,iamc,ib4}.

With respect to the new outstanding concepts introduced in \cite{gandjpa} and \cite{ib6,ib8}, some results have been communicated in the recent literature. 

For instance, weak self-adjointness of a porous medium equation and for Hamilton-Jacobi-Bellman equation were reported by Gandarias, Redondo and Bruzón \cite{gandc} and Gandarias \cite{gandna}, respectively. In a recent communication \cite{ijpa} nonlinearly self-adjoint equations of fifth-order was obtained. 

Influenced by the terminology employed by Torrisi and Tracinà in \cite{tori}, in this paper it is presented a {\it nonlinearly self-adjoint classification} of the following equation
\bb\label{1.6}
u_{t}+d(t)u_{xxxxx}+a(t)uu_{xxx}+b(t)u_{x}u_{xx}+c(t)u^{2}u_{x}=0.
\ee

Equation (\ref{1.6}) includes a large number of equations employed in Mathematical Physics such as Kaup-Kupershmidt equation, Sawada-Koterra equation, Lax equation, Caudrey-Dodd-Gibbon equation and so on. We recommend the interested reader to consult references \cite{bi,go,re,qu,ya,you,yu} and references therein to find applications of these equations in Science.

The paper deals with equation (\ref{1.1}) with the following restrictions: $(a,b,d)\neq(0,0,0)$, since the case $a=b=d=0$ is reduced to the inviscid Burgers equation and its conservation laws using the Ibragimov's developments \cite{ib1,ib2,ib3,ib6} were discussed in \cite{ijnmp, iaaca,icam}. 

The paper is organized as the follows. In the next section we determine the class of nonlinear self-adjoint equations of the type (\ref{1.6}). Next, in the section \ref{cl}, we establish conservation laws for some particular cases of (\ref{1.1}) using theorem \ref{new}.

\section{Nonlinear self-adjointness classification of equation (\ref{1.1})}\label{non}

\begin{definition}
A locally analytic function of a finite number of the variables $x,\,u$ and $u$ derivatives is called a differential function. The highest order of derivatives appearing in the differential function is called the order of this function. The vector space of all differential functions of finite order is denoted by ${\cal A}$.
\end{definition}

\begin{definition}\label{ganddef}
Equation $(\ref{1.2})$ is said to be nonlinear self-adjoint if the equation obtained from the adjoint equation $(\ref{1.3})$
by the substitution $v=\phi(x,u)$ with a certain function $\phi(x,u)\neq0$ 
is identical with the original equation $(\ref{1.2})$, that is,
\bb\label{2.4}
\left.F^{\ast}\right|_{v=\phi}=\lambda(x,u,\cdots)F,
\ee
for some $\lambda\in{\cal A}$.

Whenever $(\ref{2.4})$ holds for a certain function $\phi$ such that $\phi_{u}\neq0$ and $\phi_{x^{i}}\neq0$, equation $(\ref{1.3})$ is called weak self-adjoint.

Whenever $(\ref{2.4})$ holds for a certain function $\phi$ such that $\phi=\phi(u),\,\phi'(u)\neq0,\,\phi\neq u$, equation $(\ref{1.3})$ is called quasi-self-adjoint.

Whenever $(\ref{2.4})$ holds for $\phi= u$, equation $(\ref{1.3})$ is called (strictelly) self-adjoint.
\end{definition}

Let us determine the nonlinearly self-adjoint subclasses of the equation (\ref{1.6}). 

Let
\bb\label{2.1.1}
F=u_{t}+d(t)u_{xxxxx}+a(t)uu_{xxx}+b(t)u_{x}u_{xx}+c(t)u^{2}u_{x}.
\ee

In what follows, for simplicity, we shall write $a,\,b,\,c,\,d$ instead of $a(t)$, etc. Substituting (\ref{2.1.1}) into (\ref{1.3}) we obtain the adjoint equation $F^{\ast}=0$ to (\ref{1.6}), where
\bb\label{2.1.2}
F^{\ast}=-v_{t}+[(b-3a)u_{xx}-cu^{2}]v_{x}+(b-3a)u_{x}v_{xx}-au v_{xxx}-dv_{xxxxx}.
\ee

Thus
$$
\ba{lcl}
\left.F\right|_{v=\phi(x,t,u)}&=&\ds{-\phi_{t}-cu^{2}\phi_{x}-au\phi_{xxx}-d\phi_{xxxxx}-\phi_{u}u_{t}}\\
\\
&&\ds{+[-cu^{2}\phi_{u}+(b-3a)\phi_{xx}-3a\phi_{xxu}-5d\phi_{xxxxu}]u_{x}}\\
\\
&&\ds{+[2(b-3a)\phi_{xu}-3au\phi_{xuu}-10d\phi_{xxxuu}]u_{x}^{2}}\\
\\
&&\ds{+[(b-3a)\phi_{uu}-a\phi_{uuu}-10d\phi_{xxuuu}]u_{x}^{3}}\\
\\
&&\ds{-5d\phi_{xuuuu}u_{x}^{4}-d\phi_{uuuuu}u_{x}^{5}+[(b-3a)\phi_{x}-3au\phi_{xu}-10d\phi_{xxxu}]u_{xx}}\\
\\
&&\ds{+[2(b-3a)\phi_{u}-3au\phi_{uu}-30d\phi_{xxuu}]u_{x}u_{xx}-30d\phi_{xuuu}u_{x}^{2}u_{xx}}\\
\\
&&\ds{-10d\phi_{uuuu}u_{x}^{3}u_{xx}-15d\phi_{xuu}u_{xx}^{2}-15d\phi_{uuu}u_{x}u_{xx}^{2}}\\
\\
&&\ds{+(-au\phi_{u}-10d\phi_{xxu})u_{xxx}-20d\phi_{xuu}u_{x}u_{xxx}-10d\phi_{uuu}u_{x}^{2}u_{xxx}}\\
\\
&&\ds{-10d\phi_{uu}u_{xx}u_{xxx}-5d\phi_{xu}u_{xxxx}-5d\phi_{uu}u_{x}u_{xxxx}-d\phi_{u}u_{xxxxx}}.
\ea
$$
Assume that $\left.F\right|_{v=\phi(x,t,u)}=\lambda(x,t,u,\cdots)F$, for a certain function $\lambda\in{\cal A}$, where $F$ is given by (\ref{2.1.1}). From the $u_{t}$ coefficient it is obtained that $\lambda=-\phi_{u}$ and from the $1,\,u_{x}^{2},\,u_{xx},\,u_{x}u_{xx},\,u_{xxxx},\,u_{x}u_{xxxx}$ coefficients we conclude, respectively, that
\bb\label{2.1.3}
\ba{l}
\phi_{t}+d\phi_{xxxxx}+au\phi_{xxx}+cu^{2}\phi_{x}=0,\,\,\,\,
2(b-3a)\phi_{xu}-3au\phi_{xuu}-10d\phi_{xxxuu}=0,\\
\\
(b-3a)\phi_{x}-3au\phi_{xu}-10d\phi_{xxxu}=0,\,\,\,\,
3(b-2a)\phi_{u}-3au\phi_{uu}-30d\phi_{xxuu}=0,\\
\\
d\phi_{xu}=d\phi_{uu}=0.
\ea
\ee

From the last two equations of (\ref{2.1.3}) it is necessary to consider the cases $d=0$ and $d\neq0$.

On one hand, assuming $d=0$, we must solve the system
$$
\ba{l}
\phi_{t}+au\phi_{xxx}+cu^{2}\phi_{x}=0,\,\,\,\,
2(b-3a)\phi_{xu}-3au\phi_{xuu}=0,\\
\\
(b-3a)\phi_{x}-3au\phi_{xu}=0,\,\,\,\,
3(b-2a)\phi_{u}-3au\phi_{uu}=0.
\ea$$

On the other hand, whenever $d\neq0$, from the last two equations of (\ref{2.1.3}) we conclude that $\phi_{xu}=\phi_{uu}=0$ and the remaining equations become
$$
\ba{l}
\phi_{t}+d\phi_{xxxxx}+au\phi_{xxx}+cu^{2}\phi_{x}=0,\,\,\,\,
(b-2a)\phi_{u}=(b-3a)\phi_{x}=0.
\ea
$$

In order to not increase the volume of this paper we omit the tedious necessary calculations to obtain the solution of (\ref{2.1.3}). In the next tables it is summarized the results providing the solution of (\ref{2.1.3}). Since we have obtained many different classes of equation we shall to refer to the discovered equations as type 5-I, $\cdots$, 5-V, for equations of fifth-order, and equations type 3-I, $\cdots$, 3-IV for third-order. 

We present the conditions that the functions $a,\,b,\,c$ should satisfy in order to have nonlinearly self-adjointness, the corresponding nonlinearly self-adjoint equation and the function $\phi$. The symbol $\forall$ means that the corresponding function has no restrictions. In what follows $c_{i},\,1\leq i\leq5$, are arbitrary constants.

\begin{table}[h]\label{tab1}
\begin{tabular}{|c|c|c|c|c|c|}\hline$b$ & $c$  & Equation & $\phi$& Type\\\hline

$3a$ & $\neq0$ & $u_{t}+auu_{xxx}+3au_{x}u_{xx}+cu^{2}u_{x}=0$ & $c_{1}$& 3-I\\\hline

$3a$ & $0$ & $u_{t}+auu_{xxx}+3au_{x}u_{xx}=0$ & $c_{1}x^{2}+c_{2}x+c_{3}$& 3-II\\\hline

$0$ & $\neq0$ & $u_{t}+auu_{xxx}+cu^{2}u_{x}=0$ & $\ds{\f{c_{1}}{u}+c_{2}}$& 3-III\\\hline

$0$ & $0$ & $u_{t}+auu_{xxx}=0$ & $\ba{l}\ds{c_{1}\left(\f{x^{3}}{u}-6\int^{t} a(\tau)d\tau\right)+c_{2}\f{x^{2}}{u}}\\\ds{+c_{3}\f{x}{u}+\f{c_{4}}{u}+c_{5}}\ea$& 3-IV\\\hline
\end{tabular}
\caption{Nonlinearly self-adjoint equations of third-order ($d=0,\,a\neq0$).}
\end{table}

\begin{table}[h]\label{t2}
\begin{tabular}{|c|c|c|c|c|c|c|}\hline
$b$ & $a$ &  c  & Equation & $\phi$ & Type\\\hline

$\neq2a,\,3a$ & $\forall$ & $\forall$ & $u_{t}+du_{xxxxx}+auu_{xxx}+3au_{x}u_{xx}+cu^{2}u_{x}=0$ & $c_{1}$ & 5-I\\\hline

$3a$ & $\neq0$ & $\neq 0$ & $u_{t}+du_{xxxxx}+auu_{xxx}+3au_{x}u_{xx}+cu^{2}u_{x}=0$ & $c_{1}$ & 5-II\\\hline

$3a$ & $\neq0$ & $0$ & $u_{t}+du_{xxxxx}+auu_{xxx}+3au_{x}u_{xx}=0$ & $c_{1}x^{2}+c_{2}x+c_{3}$ & 5-III\\\hline

$2a$ & $\neq0$ & $\forall$ & $u_{t}+du_{xxxxx}+auu_{xxx}+2au_{x}u_{xx}+cu^{2}u_{x}=0$ & $c_{1}u+c_{2}$ & 5-IV\\\hline

$0$ & $0$ & $\neq0$ & $u_{t}+du_{xxxxx}+cu^{2}u_{x}=0$ & $c_{1}u+c_{2}$& 5-V\\\hline
\end{tabular}
\caption{Nonlinearly self-adjoint equations of fifth-order ($d\neq0$).}
\end{table}

{\bf Remark}: In addition to the cases listed in Table 1 and Table 2 we only have a second order nonlinear self-adjoint equation of the type (\ref{1.6}). Namely we have the equation $u_{t}+bu_{x}u_{xx}+cu^{2}u_{x}=0,$ whose corresponding $\phi$ is $\phi=const$.

The general Lax equation can be obtained from equations Type 5-III setting $d=1,\,c=3\gamma^{2}/10$ and $a=\gamma$, where $\gamma$ is real number, see \cite{bi,go}. The Ito equation ($d=1,\,a=3,\,c=2$, see \cite{bi}) belongs to the same class. The simplified modified Kawahara equation is an equation Type 5-IV, see \cite{ijpa}.

\section{Conservation laws for equation (\ref{1.6})}\label{cl}

Let
$$X=\tau(x,t,u)\f{\p}{\p t}+\xi(x,t,u)\f{\p}{\p x}+\eta(x,t,u)\f{\p}{\p u}$$
be any Lie point symmetry of (\ref{2.1.1}) and consider the formal Lagrangean given by
$${\cal L}=v[u_{t}+d(t)u_{xxxxx}+a(t)uu_{xxx}+b(t)u_{x}u_{xx}+c(t)u^{2}u_{x}].$$

Then
\bb\label{sex}
\ba{lcl}
F:=\ds{\f{\p}{\p v}{\cal L}=u_{t}+d(t)u_{xxxxx}+a(t)uu_{xxx}+b(t)u_{x}u_{xx}+c(t)u^{2}u_{x}=0},\\
\\
F^{\ast}:=\ds{\f{\p}{\p u}{\cal L}=-v_{t}+[(b-3a)u_{xx}-cu^{2}]v_{x}+(b-3a)u_{x}v_{xx}-au v_{xxx}-dv_{xxxxx}=0}.
\ea
\ee

Note that the formal Lagrangean does not depende on mixed derivatives of $u$, as well as of $u_{xxxx}$ or $u$ derivatives of order greater than $5$. From the Ibragimov's theorem on conservation laws and taking into account the last observation, the components of the conserved vector (\ref{1.4}) for the system (\ref{sex}) (observe that the first equation is (\ref{1.6})) are given by
\bb\label{3.1}
\ba{lcl}
C^{0}&=&\ds{\tau {\cal L}+W\,\f{\p {\cal L}}{\p u_{t}}},\\
\\
C^{1}&=&\ds{\xi {\cal L}+W\left[\f{\p {\cal L}}{\p u_{x}}-D_{x}\f{\p {\cal L}}{\p u_{xx}}+D_{x}^{2}\f{\p {\cal L}}{\p u_{xxx}}-D_{x}\f{\p {\cal L}}{\p u_{xxxx}}+D_{x}^{4}\f{\p {\cal L}}{\p u_{xxxxx}}\right]}\\
\\
&&\ds{+D_{x}(W)\left[\f{\p {\cal L}}{\p u_{xx}}-D_{x}\f{\p {\cal L}}{\p u_{xxx}}+D_{x}^{2}\f{\p {\cal L}}{\p u_{xxxx}}-D_{x}^{3}\f{\p{\cal L}}{\p u_{xxxxx}}\right]}\\
\\
&&\ds{+D_{x}^{2}(W)\left[\f{\p {\cal L}}{\p u_{xxx}}-D_{x}\f{\p {\cal L}}{\p u_{xxxx}}+D_{x}^{2}\f{\p {\cal L}}{\p u_{xxxxx}}\right]}\\
\\
&&\ds{+D_{x}^{3}(W)\left[\f{\p{\cal L}}{\p u_{xxxx}}-D_{x}\f{\p {\cal L}}{\p u_{xxxxx}}\right]+D_{x}^{4}(W)\f{\p{\cal L}}{\p u_{xxxxx}}}.
\ea
\ee

Observe that for the nonlinearly self-adjoint equations listed in the tables 1 and 2 we can eliminate the non-physical variable $v$ upon the substitutions given by the penultimate corresponding column of the tables. In this way we can construct conserved vectors for the equation under consideration.

\subsection{Conservation laws for equations Type 3-III}

Under the change $t\mapsto\int^{t}a(\tau)d\tau$ the class of equations 3-III is equivalent to the equation
\bb\label{3.1.1}
u_{t}+uu_{xxx}+f(t)u^{2}u_{x}=0,
\ee
for certain function $f=f(t)$. Concerning the particular case whenever $f(t)=t$, a direct inspection shows that 
\bb\label{3.1.2}
X=t\f{\p}{\p t}-u\f{\p}{\p u}
\ee
is a Lie point symmetry generator for equation
\bb\label{3.1.3}
u_{t}+uu_{xxx}+tu^{2}u_{x}=0.
\ee

Substituting $\xi=0,\,\tau=t,\,\eta=-u$ into the components (\ref{3.1}), we get
\bb\label{3.1.4}
\ba{lcl}
C^{0}&=&(tuu_{xxx}+tu^{2}u_{x}-u)v,\\
\\
C^{1}&=&-(tu^{3}+tu^{2}u_{t}+2uu_{xx}+tu_{t}u_{xx}-u_{x}^{2}-tu_{x}u_{xt}-tu_{xxt})v\\
\\
&&-(2uu_{x}+2tu_{t}u_{x}-uu_{x}-tuu_{xt})v_{x}-(u+tu_{t})uv_{xx}.
\ea
\ee

Let
$$
\ba{lcl}
\tilde{A}^{0}:&=&\left.C^{0}\right|_{v=1}=\ds{-u+D_{x}\left(\f{t^{2}u^{3}}{3}+tuu_{xx}-\f{tu_{x}^{2}}{2}\right)},\\
\\
\tilde{A}^{1}:&=&\left.C^{1}\right|_{v=1}=\ds{-tu^{3}-2uu_{xx}+\f{u_{x}^{2}}{2}-D_{x}\left(\f{t^{2}u^{3}}{3}+tuu_{xx}-\f{tu_{x}^{2}}{2}\right)}
\ea
$$
Since 
$$D_{t}\tilde{A}^{0}+D_{x}\tilde{A}^{1}=D_{t}(-u)+D_{x}\left(-tu^{3}-2uu_{xx}+\f{u_{x}^{2}}{2}\right)$$
we can take $A=(A^{0},A^{1})$, where
$$A^{0}=u,\,\,\,\,A^{1}=tu^{3}+2uu_{xx}-\f{u_{x}^{2}}{2},$$
as the conserved vector associated with the Lie point symmetry generator (\ref{3.1.2}) under the substitution $v=1$.
\subsection{Conservation laws for equations Type 3-IV}

Under the change $t\mapsto\int^{t}a(\tau)d\tau$ the class of equations 3-IV become the single equation
\bb\label{3.2.1}
u_{t}+uu_{xxx}=0.
\ee
It is clear that $(\ref{3.2.1})$ is invariant under translations in $x$. Taking $\xi=1,\,\tau=\eta=0$, the components (\ref{3.1}) become
\bb\label{3.2.2}
C^{0}=-vu_{x},\,\,\,\,
C^{1}=vu_{t}+uu_{xx}v_{x}-uu_{x}v_{xx}-2u_{x}^{2}v_{x}.
\ee
Both substitutions $v=1$ or $v=1/u$ into the components (\ref{3.2.2}) provides the trivial components $C^{0}=C^{1}=0$. However, setting $v=x/u$ into (\ref{3.2.2}) we get
$$
\ba{lcl}
A^{0}:=\ds{\left.C^{0}\right|_{v=\f{x}{u}}}&=&\ds{-\f{xu_{x}}{u}=-\ln{u}-D_{x}(x\ln{u})},\\
\\
A^{1}:=\ds{\left.C^{1}\right|_{v=\f{x}{u}}}&=&\ds{\f{xu_{t}}{u}+u_{xx}=D_{x}(x\ln{u})+u_{xx}}.
\ea
$$

Since
$$D_{t}A^{0}+D_{x}A^{1}=D_{t}(-\ln{u})+D_{x}(u_{xx})$$
we can take $A^{0}=-\ln{u}$ and $A^{1}=u_{xx}$ as components of a conserved vector for (\ref{3.2.1}).

Now consider (\ref{3.2.2}) with the substitution $v=x^{3}/u-6t$. After reckoning, the components $C^{0}$ and $C^{1}$ obtained from (\ref{3.2.2}) are
$$C^{0}=3x^{2}\ln{u}+D_{x}(6tu-x^{3}\ln{u}),\,\,\,\,C^{1}=6u-6xu_{x}+3x^{2}u_{xx}+D_{t}(x^{3}\ln{u}-6tu).$$

By transfering the terms $D_{x}(\cdots)$ from $C^{0}$ to $C^{1}$, we conclude that
$$C^{0}=3x^{2}\ln{u},\,\,\,\,C^{1}=6u-6xu_{x}+3x^{2}u_{xx}$$
provides a conservation law $D_{t}C^{0}+D_{x}C^{1}=0$ for (\ref{3.2.1}) associated with translational invariance in $x$.

\subsection{Conservation laws for equations Type 5-V}
Using the change $t\mapsto\int^{t}d(\tau)d\tau$ the class of equations 5-V become 
$$
u_{t}+u_{xxxxx}+f(t)u^{2}u_{x}=0,
$$
which is a time-dependent simplified modified Kawahara equation. Consider the particular case whenever $f(t)=u^{p}$, that is, consider the equation
\bb\label{3.3.1}
u_{t}+u_{xxxxx}+t^{p}u^{2}u_{x}=0.
\ee

A Lie point symmetry generator of (\ref{3.3.1}) is
\bb\label{3.3.2}
X=10t\f{\p}{\p t}+2x\f{\p}{\p x}-(4+5p)u\f{\p}{\p u}.
\ee

Substituting ${\cal L}=v(u_{t}+u_{xxxxx}+t^{p}u^{2}u_{x})$, $\xi=2x,\,\tau=10t$, $\eta=-(4+5p)u$, setting $v=1$ and after reckoning, it is obtained
$$
\ba{lcl}
C^{0}&=&-[(5p+4)u+2xu_{x}-10t^{p+1}u^{2}u_{x}-10tu_{xxxx}]\\
\\
&=&\ds{-(5p+2)u+D_{x}\left(10tu_{xxxx}+\f{10}{3}t^{p+1}u^{3}-2xu\right)},\\
\\
C^{1}&=&-(5p+4)t^{p}u^{3}+(2x-10t^{p+1}u^{2}u_{t}-(5p+12)u_{xxxx}-10tu_{xxxxt}\\
\\
&=&\ds{-\f{5p+2}{3}t^{p}u^{3}-(5p+2)u_{xxxx}-D_{t}\left(10tu_{xxxx}+\f{10}{3}t^{p+1}u^{3}-2xu\right)}.
\ea
$$

After transfering the term $D_{t}(\cdots)$ from $C^{1}$ to $C^{0}$ and changing the signal, the conserved vector $C=(C^{0},C^{1})$ is obtained, where
\bb\label{3.3.4}
C^{0}=(5p+2)u,\,\,\,\,
C^{1}=\ds{\f{5p+2}{3}t^{p}u^{3}+(5p+2)u_{xxxx}}.
\ee

It is interesting to observe that whenever $p=-2/5$ the conserved vector associated with the Lie point symmetry generator (\ref{3.3.2}) and upon the substitution $v=1$ is trivial. For $p=0$ we obtain a conservation law for the simplified modified Kawahara equation, see \cite{ijpa}. (Unfortunately there is a misprint in the component $C^{1}$ of the conserved vector obtained in \cite{ijpa}. The constant $5/3$ should be replaced by $1/3$. A note correcting this misprint was sent to the journal.)

\section{Conclusion}

New classes of nonlinear self-adjoint equations of the type (\ref{1.6}) were found. Namely we found four nonlinear self-adjoint equations of third-order, summarized in the Table 1. The subclasses of fifth-order nonlinear self-adjoint equations of the type (\ref{1.6}) are given in the Table \ref{t2}. A class of second-order nonlinear self-adjoint equations of the type (\ref{1.6}) was also obtained, namely, we found
$$u_{t}+bu_{x}u_{xx}+cu^{2}u_{x}=0.$$

It was showed that the class 3-IV is equivalent to the single equation (\ref{3.2.1}). Hence it was established conservation laws associated with translation in $x$, which corresponds to a momentum conservation.

The author has not found in the literature works dealing with either Lie point symmetries or conservation laws for all equations listed in the tables 1 and \ref{t2}. So it could be theme of further research to find Lie point symmetries of equations listed and, by using the functions $\phi$ and the Ibragimov's theorem, establish conservation laws for such equations.


\end{document}